# OVO FINTECH APPLICATION ANALYSIS USING THE SYSTEM USABILITY SCALE

**Luh Yuliani Purnama Dewi [1], Leon Andretti Abdillah [1,2,3,4*]**

[1] Information Systems Department, Universitas Bina Darma, Indonesia
[2] Enterprise Systems Group of Research (ES-GoR), Research Center of ICT Innovation Smart Systems & Data Science, Universitas Bina Darma, Indonesia
[3] Research Fellow, INTI International University, Malaysia
[4] Research Fellow, Chung Hua University, Taiwan
[1,2] Jalan Jenderal Ahmad Yani No. 3, Palembang, Indonesia
Sur-el : luhyulianidewi93@gmail.com [1] , leon.abdillah@yahoo.com [2]
[*)] **Corresponden Author**
DOI : https://doi.org/10.33557/jurnalmatrik.v26i2.3331

***Abstract:*** *The advancement of information technology has propelled payment systems from conventional methods to technology-based solutions, such as e-wallets and Financial Technology (Fintech). Fintech, a fusion of technology and financial services, has evolved into an online business model enabling fast and remote transactions. This research discusses the progress of information technology (IT) influencing payment systems, particularly in the realm of Fintech. The primary focus is on the Fintech application OVO and its impact on tenants at the International Plaza (IP) Mall in Palembang. This study employs the System Usability Scale (SUS) to evaluate the Usability of the OVO application, emphasizing aspects like effectiveness, efficiency, and user satisfaction. The research is descriptive and quantitative, with a sample of 50 respondents from Mall IP tenants. Data is collected through SUS questionnaires and analyzed using SPSS. The evaluation indicates that the OVO application has high Usability , with an SUS score of 87.05 (Grade A), signifying an "Excellent" rating. It suggests that the OVO application provides a comfortable user experience, particularly in electronic financial transactions.*

***Abstrak:*** *Kemajuan teknologi informasi telah mendorong sistem pembayaran dari metode konvensional ke solusi berbasis teknologi, seperti dompet elektronik dan Teknologi Finansial (Fintech). Fintech, perpaduan teknologi dan layanan keuangan, telah berevolusi menjadi model bisnis daring yang memungkinkan transaksi cepat dan jarak jauh. Penelitian ini membahas kemajuan teknologi informasi (TI) yang memengaruhi sistem pembayaran, khususnya di bidang Fintech. Fokus utama adalah pada aplikasi Fintech OVO dan dampaknya terhadap penyewa di Mal International Plaza (IP) di Palembang. Penelitian ini menggunakan Skala Kegunaan Sistem (SUS) untuk mengevaluasi Kegunaan aplikasi OVO, dengan menekankan aspek-aspek seperti efektivitas, efisiensi, dan kepuasan pengguna. Penelitian ini bersifat deskriptif dan kuantitatif, dengan sampel 50 responden dari penyewa IP Mall. Data dikumpulkan melalui kuesioner SUS dan dianalisis menggunakan SPSS. Evaluasi menunjukkan bahwa aplikasi OVO memiliki Kegunaan yang tinggi, dengan skor SUS 87,05 (Nilai A), yang menandakan peringkat "Sangat Baik". Hal ini menunjukkan bahwa aplikasi OVO memberikan pengalaman pengguna yang nyaman, terutama dalam transaksi keuangan elektronik.*

***Kata-kata kunci:*** *Fintech, International Plaza, OVO, SUS, Tenant.*

## 1. INTRODUCTION

Along with advances in information technology (IT), payment technology has also experienced rapid progress. Previously, the payment system was carried out using conventional methods. So, in electronic commerce (e-commerce) mode, the payment system uses information technology such as e-wallet, e-payments, financial technology (Fintech), and so on [1]. The ability of information technology to make it easy to store data quickly and efficiently and to access it anywhere and anytime has given rise to the financial technology or Fintech industry [2].

Financial technology (Fintech) is a combination of technology with financial/financial services which has finally developed towards a business model from conventional to online, where initially you had to meet face to face to pay and bring a certain amount of cash, now you can make long distance transactions by making payments that can be made done in just seconds [3]. The use of Fintech can maximize banking services to consumers. Problems in buying and selling transactions and payments include not having time to look for goods at shopping places, going to the bank/ATM to transfer funds, customers' reluctance to visit places where the service is less than pleasant. Fintech helps buying and selling transactions so that the payment system becomes more efficient and economical but remains effective.

OVO is one of the unicorn Fintech companies in Indonesia [4] and is widely used to support transactions for various business activities, especially in urban areas. OVO is a popular fintech application in Indonesia that provides a variety of financial services, including digital payments and investment management. OVO's user-friendly design and different features make it simple for users to do daily operations such as bill payments, shopping, and cash transfers.

Usability is a very important thing in interaction design which includes: behavior, efficiency, effectiveness, flexibility, security, utility, ease of learning, and ease of remembering [5]. Usability is defined as the extent to which a program can be used to achieve measurable goals with effectiveness, efficiency, and satisfaction in a particular context of use [6].

According to the System Usability Scale (SUS), OVO may be evaluated in terms of simplicity of use, user happiness, and efficiency in reaching financial goals, confirming its position as one of the most innovative and popular digital financial solutions among the general population.

The OVO Fintech Application Analysis The System Usability Scale (SUS) is used to evaluate the usability of the OVO app, a major participant in Indonesia's digital wallet sector. SUS, a well-known tool developed by John Brooke in 1986, has been used in studies to get insight on the application's user experience and functionality. In a study of e-wallet apps such OVO, Dana, and ShopeePay, researchers used SUS, UX Honeycomb, and UEQ to measure usability levels. These investigations found that while OVO scores highly in terms of usefulness and value, its usability falls short, notably in areas relating to efficiency and reliability, rating lower than its competitors such as DANA and Go-Pay [7]. Another research focused on OVO and utilized UEQ to assess user experience factors such as efficiency, perspicuity, dependency, stimulation, attractiveness, and novelty. OVO excels in efficiency and dependability, but lacks fresh experiences and appeal, resulting in a poor rating on Google PlayStore [8]. Evaluations comparing OVO to other popular e-wallets revealed numerous mistakes and lowered user satisfaction, underlining the need for continual development in system usability [9]–[11]. This research emphasizes the need of improving both functional dependability and innovative appeal inside the OVO app architecture in order to improve overall usability and user happiness.

## 2. RESEARCH METHODOLOGY

This section is divided into five core subsections, each describing an important part of the research process. The first paragraph, Research Design, describes the study's broad methodology and structure, with an emphasis on evaluating the usability of the OVO fintech application. The second subsection, Participants, describes the selection process for participants as well as the user sample's demographics. The third area, Data Collection, discusses the methods and tools used to collect usability data, such as task-based testing and the SUS questionnaire. The fourth component, Data Analysis, describes the methodologies used to process and evaluate usability data. Finally, the fifth subsection, Tools and Technologies, discusses the software and systems utilized to perform the study and analysis.

### 2.1 Research Design

This research method is quantitative descriptive. Quantitative research is research by obtaining data in the form of numbers or qualitative data that is numeric [12], [13]. The research strategy used in this study is quantitative, with the System Usability Scale (SUS) technique used to assess the usability of the OVO app. The SUS [14] assesses the perceived usability of systems such as software and websites. Participants scored a 10-item questionnaire on a 5-point Likert scale, indicating their level of agreement with statements about the application's ease of use and usefulness. This systematic technique allows for statistical analysis of user satisfaction, resulting in trustworthy data that can be used to evaluate the application's effectiveness in real-world circumstances.

The SUS approach is especially appreciated for its efficiency and validity in usability testing. It has been verified via substantial research including thousands of surveys across multiple studies [15]. Researchers can acquire meaningful information from users in many digital situations, making it a favored alternative for usability studies [16]. The SUS questionnaire's alternating positive and negative items reduce acquiescence bias and ensure actual user impressions, rather than an inclination to agree with assertions. Overall, the SUS offers a solid framework for assessing user experience, making it an indispensable tool for developers and academics looking to improve program usability.

## 2.2   Participants

Participants in this study were OVO application users, especially tenants at the International Plaza (IP) Mall in Palembang City. The total number of participants who participated was 50 respondents. The data obtained were the results of Usability measurements using the System Usability Scale (SUS) method. This entailed distributing questionnaires both directly and through Google Forms, allowing for a large reach and rapid data collecting. Such approaches are important in quantitative descriptive research because they allow for systematic examination of user experiences and preferences, providing to a greater knowledge of financial application usability in real-world scenarios [17], [18]. The study uses these methodologies to obtain significant information for future OVO platform developments.

## 2.3   Data Collection

This study uses two data gathering methods: primary data and secondary data. Primary data is collected directly from people via usability testing and the System Usability Scale (SUS) questionnaire, and secondary data is sourced from current literature, research publications, and pertinent documentation on the OVO fintech application.

Primary data is data collected directly by researchers from original sources. For this study, primary data was obtained from direct observation, interviews, and questionnaires. The stages of primary data collection include: First, observation: Researchers directly observed the location at Mall International Plaza (IP) Palembang to collect data and information related to the use of the OVO fintech application by tenants. Second, interviews: Researchers conducted direct interviews with tenant managers who used OVO fintech at Mall International Plaza (IP) Palembang. Third: questionnaires: Researchers distributed questionnaires to predetermined respondents, containing questions according to the System Usability Scale (SUS) method.

Meanwhile, secondary data is obtained from literature studies that include reading manuals, papers, journals, and other reports relevant to the research objectives. This method involves collecting journals, literature, papers, books, and sources from the internet as references for research.

## 2.4   Tools and Technologies

The OVO fintech application was analyzed using the System Usability Scale (SUS), a powerful method for assessing usability across several digital platforms. The SUS (Figure 1) is a 10-item questionnaire with a 5-point Likert scale for collecting and analyzing user satisfaction and application effectiveness [19].

Microsoft Excel is an efficient tool for determining System Usability Scale (SUS) scores for analyzing the OVO fintech application. By entering user replies into Excel, researchers may quickly use the SUS scoring method, which adds adjusted scores from odd and even questions and multiplies the result by 2.5. This technique improves analysis efficiency by streamlining data processing and enabling speedy computation and presentation of usability measures. Furthermore, using automated scripts in Excel may greatly minimize human mistakes and enhance productivity, making it a solid alternative for usability studies [20]

**Figure 1. SUS Questionnaire**

### 2.5 Data Analysis

Data Analysis computed the SUS scores by adding the corrected responses from ten surveys and multiplying by 2.5 to get a final score of 100. A score of 70 indicates good usability, consistent with earlier research indicating that higher ratings lead to better user experiences [14]. SUS ratings have a strong association with user satisfaction indicators, indicating their usefulness in evaluating financial apps [21], [22].

Users will have scored the ten template questions (Figure 1) on a scale of 1 to 5, indicating their level of agreement. The calculation steps: 1) Subtract one point from your score for each odd-numbered question, 2) Subtract 5 from each even-numbered question, and 3) Take the new values you discovered and add them to the overall score. Then double it by 2.5.

To compute the SUS score [23], add the score contributions of each item. Each item's score contribution will be between 0 and 4. For items 1, 3, 5, 7, and 9, the score contribution equals the scale position minus one. For items 2, 4, 6, 8, and 10, the contribution is 5 less the scale position. To get SU's overall worth, multiply the total of the scores by 2.5 [21]. SUS scores range from 0 to 100. All of these complex calculations have resulted in the total score of 100. This is not a percentage, but it is an easy method to view your score.

$$
\begin{aligned}
SUS\ Score\ = &((Q1-1) + (Q3-1) \\
&+ (Q5-1) + (Q7-1) \\
&+ (Q9-1) + (5-Q2) \\
&+ (5-Q4) + (5-Q6) \\
&+ (5-Q8) + (5-Q10)) \\
&* 2.5
\end{aligned}
\quad (1)
$$

The score calculation rule applies to only one responder. To get the mean SUS score for each responder, sum their full scores and divide by the number of respondents [21]. Here is the formula for calculating the SUS mean score:

$$\bar{x} = \frac{\sum x}{n} \quad (2)$$

Notes:

$\bar{x}$ = mean score

$\sum x$ = Total SUS Scores

n = Numbers of participants

SUS scores are often considered a percentage, but they are not. So it is necessary to interpret the SUS score as an Adjective Rating Scale [24] (Figure 2).

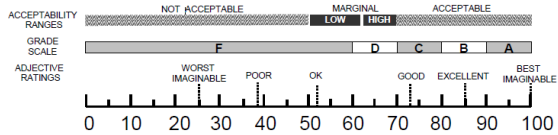

Figure 2. SUS Score Interpretation

## 3. RESULTS AND DISCUSSION

OVO is one type of financial technology (Fintech) innovation in the payment system used in online traffic or online motorcycle taxis. OVO can be used for Grab and Tokopedia services. In addition to the online motorcycle taxi program, OVO collaborates with several Indonesian MSME companies. OVO can be used to pay for purchases of goods from consumers to small and medium enterprises. OVO is a smart application that makes it easy for its users to make transactions and payments online.

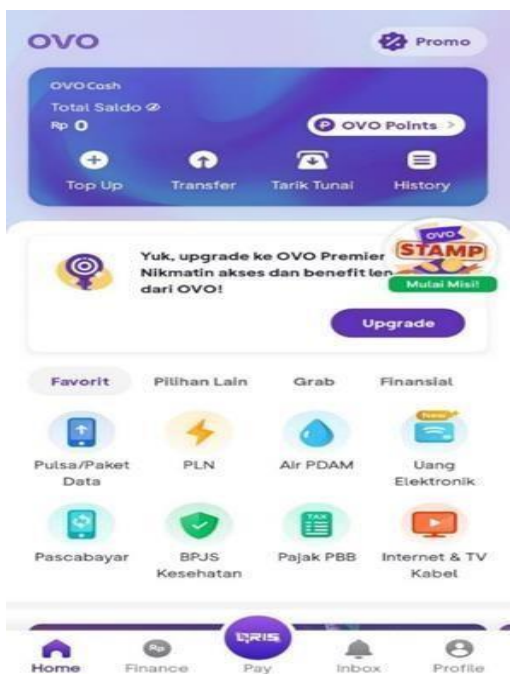

Figure 3. OVO Main Menu

Below are described one by one the superior features of OVO as follows: 1) Home: This menu will show the amount of OVO Cash and OVO Points that have been owned, in addition there are three submenus, 2) Transfer: This feature provides transfer services to fellow OVO or to bank accounts, 3) Scan: This feature provides a QR code or barcode scanning service to make payment transactions, 4) OVO ID: This feature contains the user's identity code in the form of a barcode or QR Code, 5) Deals: This feature will show various interesting current promos held by this application with its partner merchants. The promos are divided into three categories, namely Cashback Again and Again, Happiness Column, Favorites and Efficient, and the Other Worldly Delights category, 6) Finance: This feature will make it easier for those of you who want to invest with a profit of 7 percent per year with a minimum transaction of IDR 10 thousand, 7) Wallet: In this feature, you can see the purple membership card that contains the user's name and membership number, and 8) History: This feature will show all payment and top-up transactions that you have made to date, both using OVO Cash and OVO Points.

### 3.1 Participant Characteristics

The total participants involved in this study were 50, who were tenants at the Mall International Plaza Palembang. These tenants use OVO as a payment method for product and service transactions.

Based on gender, the tenant participants were dominated by female participants as many as 34 people or 68%, while the male participants numbered 16 or 32%. This indicates a big female user base. Women are rapidly engaging with digital financial services, reflecting social developments towards inclusion in fintech adoption [25]–[27].

The OVO fintech application research shows that 36% of users are Club members, while 64% are Premier members. The distribution shows a substantial preference for the Premier tier, indicating that users enjoy the improved features and perks associated with this membership level, as indicated by usability research on fintech applications [9], [25].

The investigation of OVO's fintech application showed that 64% of customers use the Rp. 20,000,000 transaction limit, while 36% use the Rp. 40,000,000 limit. This preference for smaller limitations may reflect customer worries about transaction security and usability, underlining the need of enhanced system quality in increasing user happiness and confidence in digital financial services.

**Table 1. Tenants' Income**

| No | Tenant Income Per Month (Rp.) | F | Percentage |
|---|---|---|---|
| 1 | < 1.000.0000 | 20 | 40% |
| 2 | 1.000.000 - 4.500.000 | 10 | 20% |
| 3 | 4.500.000- 10.000.000 | 7 | 14% |
| 4 | 10.000.00- 15.000.000 | 4 | 8% |
| 5 | > 15.000.000 | 9 | 18% |

The examination of renter income for OVO fintech application users demonstrates that respondents came from varied economic backgrounds. Notably, 40% of users earn less than Rp. 1,000,000, with 20% falling between Rp. 1,000,000 and Rp. 4,500,000. A smaller portion, 14%, earns between Rp. 4,500,000 and Rp. 10,000,000, while just 8% earn up to Rp. 15,000,000. Interestingly, no users reported revenues between Rp. 15,000,000 and Rp. 20,000,000, but 18% earned more than Rp. 20,000,000. This distribution (Table 1) demonstrates the application's reach across income levels and potential for financial inclusion in Indonesia [17], [25], [26], [28], [29].

Respondents' use of the OVO fintech application for an extended period of time suggests a high degree of engagement. 40% of users have used the app for less than a year, while 30% have used it for more than three years. In addition, 20% claim usage of more than two years, while 10% fall into other groups. This distribution indicates a rising user base, emphasizing the application's potential for long-term adoption and enjoyment, as indicated by usability studies in financial technology.

### 3.2  Participant Business Sectors

The product options in the OVO fintech application include a wide variety of sectors (Figure 4). 40% of consumers engage with fashion things, while 30% choose food items. Electronics interest 20% of customers, with just 10% purchasing cellphones. This distribution emphasizes the application's adaptability to varied customer preferences, highlighting its significance in improving online shopping experiences in Indonesia, as evidenced by recent research on digital wallet usage and user happiness in fintech contexts.

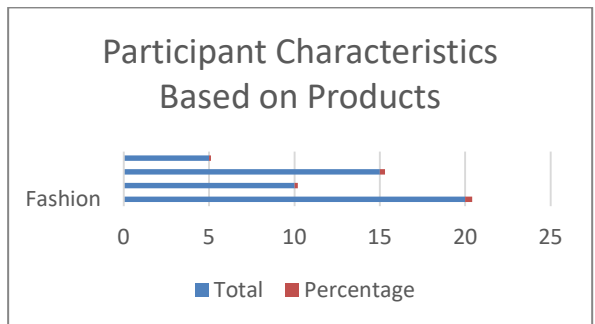

**Figure 4. Participant Characteristics Based on Products**

### 3.3 Average Score Per Question

Figure 5 depicts the average score from the ten System Usability Scale (SUS) items for the OVO fintech application, suggesting that users have a favorable opinion of usability overall. This score identifies strengths and areas for development in fintech apps, with a focus on improving user experience and resolving problems from earlier research (1, 2, 3). Such evaluations are critical for ensuring that the program effectively satisfies the demands of its users while remaining competitive in the digital wallet industry.

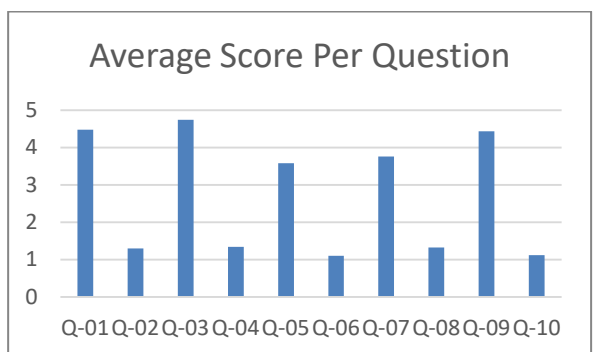

**Figure 5. Average Score Per Question**

### 3.4 SUS Score Calculation Results

A System Usability Scale (SUS) score of 87.05 shows great usability (Table 2), above the average score of 68, which is regarded the 50th percentile in usability studies [30]. This number indicates that users consider the system to be exceptional, since scores above 80.3 are classified as such. John Brooke established the SUS in 1986, which has been verified via considerable study and is frequently used to evaluate the usability of numerous systems, including websites and apps [31].

**Table 2. SUS Questionnaire Calculation**

| P | SUS | P | SUS | P | SUS | P | SUS | P | SUS |
|---|---|---|---|---|---|---|---|---|---|
| 1 | 97,5 | 11 | 90 | 21 | 80 | 31 | 85 | 41 | 82,5 |
| 2 | 92,5 | 12 | 97,5 | 22 | 77,5 | 32 | 95 | 42 | 87,5 |
| 3 | 90 | 13 | 85 | 23 | 82,5 | 33 | 87,5 | 43 | 85 |
| 4 | 92,5 | 14 | 85 | 24 | 77,5 | 34 | 87,5 | 44 | 87,5 |
| 5 | 85 | 15 | 82,5 | 25 | 77,5 | 35 | 87,5 | 45 | 85 |
| 6 | 82,5 | 16 | 87,5 | 26 | 90 | 36 | 100 | 46 | 87,5 |
| 7 | 87,5 | 17 | 87,5 | 27 | 87,5 | 37 | 95 | 47 | 80 |
| 8 | 85 | 18 | 87,5 | 28 | 90 | 38 | 90 | 48 | 82,5 |
| 9 | 92,5 | 19 | 92,5 | 29 | 80 | 39 | 97,5 | 49 | 70 |
| 10 | 87,5 | 20 | 90 | 30 | 90 | 40 | 95 | 50 | 75 |

A Net Promoter Score (NPS) score of 87.05 indicates that the system is both user-friendly and likely to be suggested by users, coinciding with the criteria of a "Promoter" [32].

### 3.5 SUS Score Interpretation

It was found that the final average score was 87.05, meaning the score was above the average of 68, this means that based on the SUS scoring table that has been discussed above, the OVO application is at Grade "A" with an "Excellent" rating predicate (see Figure 2).

## 4. CONCLUSION

Based on the results of the questionnaire that has been distributed at the international plaza mall (IP) Palembang, the following research results were obtained: 1) The results obtained from the Usability Evaluation of the OVO application using the SUS (System Usability Scale) method on 50 respondents who use the OVO application, the total SUS score calculation was 4217.5, divided by the number of respondents of 50, which resulted in 84.35. So the overall average of respondents is 84.35, this means that based on the SUS scoring table, the OVO application is at Grade A with an Excellent rating predicate, and 2) With the results of the usability evaluation on the OVO application, which is currently more than enough to make users comfortable, especially in electronic financial transactions.

The limitations of this study include still being located in one mall, analyzing one fintech application, with limited analysis using SUS. For further research, it can be expanded again by involving tenants in many malls, using more fintech applications and combined with a number of methods.

**ACKNOWLEDGEMENT**

The authors would like to express their deepest gratitude to the Center for Research on ICT Innovation, Smart Systems, and Data Science, Bina Darma University and the Management of Internasional Plaza who have provided much assistance so that this research can be carried out properly.